\documentclass[aps,twocolumn,showpacs,superscriptaddress,groupedaddress,nofootinbib]{revtex4} 
\usepackage{graphicx}
\usepackage{fancyhdr}
\usepackage{braket}
\usepackage[sort&compress]{natbib}
\usepackage{subfigure}
\usepackage{amsmath}
\usepackage{amssymb}
\usepackage{amsfonts}
\usepackage{rotating}
\usepackage{mathtools}
\usepackage{cancel}
\usepackage{lipsum}
\usepackage{mathtools}
\usepackage{color}
\usepackage{bbm}
\usepackage{dsfont}
\usepackage{bbold}
\usepackage{multirow}
\usepackage{ulem}
\newcommand{\eps}{\epsilon}
\begin{document}

\title{Off-shell ambiguities in correlation functions: strategies to minimize them}
\author{R. Molina}
\email{Raquel.Molina@ific.uv.es}
\affiliation{Departamento de F\'{\i}sica Te\'orica and IFIC,
Centro Mixto Universidad de Valencia-CSIC, Parc Científic UV, C/ Catedrático José Beltrán, 2, 46980 Paterna, Spain}   
\author{E. Oset}
\email{Eulogio.Oset@ific.uv.es}
\affiliation{Departamento de F\'{\i}sica Te\'orica and IFIC,
Centro Mixto Universidad de Valencia-CSIC, Parc Científic UV, C/ Catedrático José Beltrán, 2, 46980 Paterna, Spain}

\begin{abstract}  
We face here the problem of uncertainties in correlation functions due to the freedom in the off-shell dependence of the wave functions, or equivalently, off-shell ambiguities in the scattering matrices. We make the study for the case of meson baryon interaction, choosing the $K\Lambda$, $K \Sigma$, $\eta N$ coupled channels, which are related to the $N^*(1535)$ resonance. We find that using realistic interactions based on chiral dynamics the uncertainties are small, of the order of $2-3$~\%, but could be much bigger if other methods are used. However, our study shows the way to optimally overcome these uncertainties in the analysis of correlation functions, and we provide a series of recommendations for any general analysis.

\end{abstract}
\maketitle
\section{Introduction}
A recent paper \cite{epelulf} has pointed out to an issue that, although known, had not yet been given the necessary attention. This is related to the common assumption that the source function $S_{12}(r)$, used to evaluate correlation functions of pairs of particles, is universal, meaning that it depends on the type of high energy reaction, $pp$, $pA$, $AA$, used to produce the investigated pair, and not on the pairs produced, nor on the theoretical scheme used to analyze the correlation functions. The paper shows clearly that this is not the case and the source depends on the implicit off-shell form of the wave functions used, or equivalently, on the related off-shell dependence of the scattering amplitudes of the pair. 

We repeat here the argument used in \cite{epelulf}. The correlation function can be written as 
\begin{eqnarray}
C(\vec{k})=\braket{\psi^{(+)}_{-\vec{k}}|\hat{S}_{12}|\psi^{(+)}_{-\vec{k}}}\ ,
\end{eqnarray}
where $\psi^{(+)}_{-\vec{k}}$ is the wave function for the interaction of the observed pair and $\hat{S}_{12}$ is the operator for the source function, which approximately takes into account the probability that the pair is produced at a relative distance $r$. This source function is taken as a local operator and again, following Ref.~\cite{epelulf}, it is written as,
\begin{equation}
 \braket{\vec{r}\,'|\hat{S}_{12}|\vec{r}}=\delta(\vec{r}-\vec{r}\,')S_{12}(r)\ ,
\end{equation}
where $S_{12}(r)$ is usually taken as a Gaussian,
\begin{eqnarray}
 S_{12}(r)=\frac{e^{-\frac{r^2}{4R^2}}}{(4\pi R^2)^{3/2}}\ .~\label{eq:sc}
\end{eqnarray}
Since the correlation function $C(k)$ is an observable, it should be invariant under unitarity transformations that change the off-shell behaviour of the wave function, hence we write as in \cite{epelulf},
\begin{eqnarray}
 C(\vec{k})=\braket{\psi^{(+)}_{-\vec{k}}\hat{U}^\dagger|\hat{U}\hat{S}_{12}\hat{U}^\dagger|\hat{U}\psi^\dagger_{-\vec{k}}}\ ,
\end{eqnarray}
which means that we are free to make any unitary transformation in the wave function,  but the source should also be transformed with the same unitary transformation. This is the essence of the message in \cite{epelulf}, which is certainly unquestionable: the source function cannot be universal, since it is tied to the theoretical scheme used to study the interaction of the particles and its implicit off-shell effects, which will change from one scheme to another. Viewed from a different perspective, different schemes or theoretical models will use different Hamiltonians to produce the wave function, which will lead to the same on-shell properties, for instance phase shifts, but different off-shell behaviour of the scattering matrix. This should revert on different values obtained for the correlation function if the same source function is used, as one does by accepting the universality of the source.

An example is shown in \cite{epelulf} for the nucleon-nucleon correlation function implementing two particular unitary transformations, where it is shown that the results for the correlation functions using the same source function are indeed quite different. 

In the same work it is, however, mentioned that, realistic models of hadron interaction are constrained by some physical principles and the remaining off-shell ambiguities may therefore be expected to be less pronounced in practice, continuing as ``The above power-counting based-argument points towards a mild scheme dependence of $NN$ interaction in chiral effective theories''. Similar conclusions are obtained in \cite{kievsky}, where using the same source and very different models of the $NN$ interaction, the correlation functions obtained are very similar, with differences smaller than $5$\%. 

There is another important conclusion in \cite{epelulf}, which is that when it comes to the study of three body systems, or three body correlation functions, the off-shell effects of the two body amplitudes are much more important, and the empirical three body forces used to fit some data are highly model dependent. In this sense, we call the attention to the works of \cite{alber1,alber2} for the systems of two mesons and one baryon, or three mesons, respectively, where using chiral Lagrangians one finds that the off-shell two body amplitudes used with Fadeev equations generate three body forces, but the same chiral Lagrangians generate contact three body forces that cancel exactly the forces generated by the off-shell part of the two-body amplitudes. This is a nice feature of the chiral Lagrangians, which lead to a result where the effect of the non-physical off-shell amplitudes dissappear at the end. Yet, if a different model is used for the two body amplitudes, a different three body force would have to be introduced empirically to get the same final three body amplitudes. In summary, the three body force is not an observable, and as shown in \cite{epelulf}, it is highly model dependent. 

In what follows we look in detail into the off-shell ambiguities in the correlation functions for the case of the meson-baryon interaction, and we choose a particular case for the $K\Lambda$, $K\Sigma$, $\eta\Lambda$, $\pi N$ coupled channels interaction, which leads to the generation of the $N^*(1535)$ resonance \cite{wolfram, inoue}. The correlation functions for these systems have been recently calculated in \cite{raquexiao}. 

\section{Formalism}
We will use a formalism for the correlation functions based on scattering matrices in momentum space, following closely the known Koonin-Pratt formalism~\cite{koonin, prat, geng, valcola} (see Ref.~\cite{juancorre}) for discussion in different formalisms). For this purpose, the first step is to construct the scattering matrices. We follow the formalism of \cite{danijuan} and start with an interaction potential between two particles which has a separable form in momentum space,
\begin{eqnarray}
 V(\vec{q},\vec{q}\,')=V\theta(q_{\mathrm{max}}-|\vec{q}|)\theta(q_\mathrm{max}-|\vec{q}\,'|)\ .\label{eq:pot}
\end{eqnarray}
With this we obtain the scattering matrix which is also separable as~\cite{danijuan},
\begin{eqnarray}
 T(\vec{q},\vec{q}\,')=T\theta(q_\mathrm{max}-|\vec{q}\,|)\theta(q_\mathrm{max}-|\vec{q}\,'|)\ ,\label{eq:tm}
\end{eqnarray}
and then,
\begin{eqnarray}
 T=\frac{V}{1-VG}\label{eq:bethe}
\end{eqnarray}
with $G\equiv G(s)$, the loop function for the two particles which for meson-baryon is given by
\begin{eqnarray}
 G=\int^{q_{\mathrm{max}}}\frac{q^2dq}{(2\pi)^2}\frac{\omega_1(q)+\omega_2(q)}{2\omega_1(q)\omega_2(q)}\frac{2M}{s-(\omega_1(q)+\omega_2(q))^2+i\eps}\ ,\nonumber\\
 \label{eq:lp}
\end{eqnarray}
with $\omega_i(q)=\sqrt{q^2+m_i^2}$, $i=1,2$, $M$ the baryon mass and $s=M_{\mathrm{inv}}^2(1,2)$. Eq.~(\ref{eq:bethe}) in the case of coupled-channels is,
\begin{eqnarray}
 T=[1-VG]^{-1}V\ ,
\end{eqnarray}
with $G=\mathrm{diag}(G_i)$. Note that $V,G,T$ depend only on $M_{\mathrm{inv}}$. We observe that the loop function has been regularized with the cutoff $q_\mathrm{max}$ as a consequence of the form of the potential, Eq.~(\ref{eq:pot}). Although we have make the derivation starting from a separable potential, the result is the same as obtained in \cite{ollerulf} by using unitarity and dispersion relations, only the loop function of Eq.~(\ref{eq:lp}) is here regularized with a cutoff, while in \cite{ollerulf} dimensional regularization is used, and the approximate equivalence of both methods is also discussed in~\cite{ollerulf}. The scheme described here, together with the use of the interaction potential derived from chiral Lagrangians, is the widely used chiral unitary approach. The formulation given here has two virtues: The first one is that the cutoff $q_\mathrm{max}$ which regularizes the $G$ loop function is a measure of the range of the interaction in momentum space, as seen in Eq.~(\ref{eq:pot}). The second one is that one has an explicit structure for the on-shell and off-shell parts of the scattering matrix. Indeed, $T$ depends only on $M_\mathrm{inv}$, and this is the on-shell scattering matrix. The factor $\theta(q_\mathrm{max}-|\vec{q}\,'|)$ provides the off-shell behaviour. The separation is most welcome when it comes to test the off-shell effects of the wave function, which are given by the off-shell effects of the $T$-matrix in our formalism.
\subsection{Correlation functions}
The correlation function is given by
\begin{eqnarray}
 C(p)=\int d^3r S_{12}(r)|\psi(\vec{p},\vec{r}\,)|^2\ ,
\end{eqnarray}
where $\psi(\vec{p},\vec{r}\,)$ is the pair wave function and $S_{12}(r)$ the source function parametrized by a gaussian as shown in Eq.~(\ref{eq:sc}). In Ref.~\cite{raquexiao} the coupled channels $K^0\Sigma^+$, $K^+\Sigma^0$, $K^+\Lambda$, $\eta p$, were used ignoring the far away $\pi N$ channel, in order to investigate correlation functions for these four channels. Following the formalism of Ref.~\cite{valcola}, the correlation functions were evaluated in terms of the scattering matrices with the results,
\begin{eqnarray}\label{eq:C1}
  C_{K^0\Sigma^+} (p_{K^0})&&= 1+4\,\pi\, \theta(q_{\rm max}-p_{K^0})\, \int dr \, r^2 S_{12}(r)  \cdot \nonumber\\[2mm]
  && \left\{ \left|j_0(p_{K^0}\, r)+T_{K^0 \Sigma^+, K^0 \Sigma^+}(E)\; \tilde{G}^{(K^0\Sigma^+)}\right|^2 \right.  \nonumber\\[2mm]
  &&\;+ \left|T_{K^+ \Sigma^0, K^0 \Sigma^+}(E)\; \tilde{G}^{(K^+\Sigma^0)}\right|^2
  \nonumber\\&&\,+  \left|T_{K^+ \Lambda, K^0 \Sigma^+}(E)\; \tilde{G}^{(K^+\Lambda)} \right|^2     \nonumber\\[2mm]
  &&\;\left. +\left|T_{\eta p,  K^0 \Sigma^+}(E)\; \tilde{G}^{(\eta p)}\right|^2 - j_0^2 (p_{K^0}\, r)\right\},  \nonumber\\
\end{eqnarray}
\begin{eqnarray}\label{eq:C2}
  C_{K^+\Sigma^0} (p_{K^+})&&= 1+4\,\pi\, \theta(q_{\rm max}-p_{K^+})\, \int dr \, r^2 S_{12}(r)  \cdot \nonumber\\[2mm]
  && \left\{ \left|j_0(p_{K^+}\, r)+T_{K^+ \Sigma^0, K^+ \Sigma^0}(E)\; \tilde{G}^{(K^+\Sigma^0)}\right|^2 \right.  \nonumber\\[2mm]
  &&\;+ \left|T_{K^0 \Sigma^+, K^+ \Sigma^0}(E)\; \tilde{G}^{(K^0\Sigma^+)}\right|^2
  \nonumber\\&&\,+  \left|T_{K^+ \Lambda, K^+ \Sigma^0}(E)\; \tilde{G}^{(K^+\Lambda)} \right|^2     \nonumber\\[2mm]
  &&\;\left. +\left|T_{\eta p, K^+ \Sigma^0}(E)\; \tilde{G}^{(\eta p)}\right|^2 - j_0^2 (p_{K^+}\, r)\right\},  \nonumber\\
\end{eqnarray}
\begin{eqnarray}\label{eq:C3}
  C_{K^+\Lambda} (p_{K^+})&&= 1+4\,\pi\, \theta(q_{\rm max}-p_{K^+})\, \int dr \, r^2 S_{12}(r)  \cdot \nonumber\\[2mm]
  && \left\{ \left|j_0(p_{K^+}\, r)+T_{K^+ \Lambda, K^+ \Lambda}(E)\; \tilde{G}^{(K^+\Lambda)}\right|^2 \right.  \nonumber\\[2mm]
  &&\;+ \left|T_{K^0 \Sigma^+, K^+ \Lambda}(E)\; \tilde{G}^{(K^0\Sigma^+)}\right|^2
   \nonumber\\&&\,+  \left|T_{K^+ \Sigma^0, K^+ \Lambda}(E)\; \tilde{G}^{(K^+\Sigma^0)} \right|^2     \nonumber\\[2mm]
  &&\;\left. +\left|T_{\eta p, K^+ \Lambda}(E)\; \tilde{G}^{(\eta p)}\right|^2 - j_0^2 (p_{K^+}\, r)\right\},  \nonumber\\
\end{eqnarray}
\begin{eqnarray}\label{eq:C4}
  C_{\eta p} (p_{\eta})&&= 1+4\,\pi\, \theta(q_{\rm max}-p_{\eta})\, \int dr \, r^2 S_{12}(r)  \cdot \nonumber\\[2mm]
  && \left\{ \left|j_0(p_{\eta}\, r)+T_{\eta p, \eta p}(E)\; \tilde{G}^{(\eta p)}\right|^2 \right.  \nonumber\\[2mm]
  &&\;+ \left|T_{K^0 \Sigma^+, \eta p}(E)\; \tilde{G}^{(K^0\Sigma^+)} \right|^2
   \nonumber\\&&\,+  \left|T_{K^+ \Sigma^0, \eta p}(E)\; \tilde{G}^{(K^+\Sigma^0)}\right|^2     \nonumber\\[2mm]
  &&\;\left. +\left|T_{K^+ \Lambda, \eta p}(E)\; \tilde{G}^{(K^+ \Lambda)}\right|^2 - j_0^2 (p_{\eta}\, r)\right\},  %\nonumber\\[2mm],
\end{eqnarray}
where $p_i$ is the momentum of the particles in the rest frame of the pair,
\begin{equation}\label{eq:pi}
  p_i=\dfrac{\lambda^{1/2}(s, m_i^2, M_i^2)}{2\, \sqrt{s}},
\end{equation}
with $m_i, M_i$ being the masses of the meson, baryon of the channel considered, and the $\tilde{G}^{(i)}\equiv\tilde{G}^{(i)}(r; E)$ function is given by
\begin{equation}\label{eq:G2}
  \tilde{G}^{(i)}= \int \dfrac{{\rm d}^3 q}{(2\pi)^3} \,\; \dfrac{\omega_i(q)+E_i(q)}{2\,\omega_i(q)\, E_i(q)}\; \dfrac{2M_i \,j_0(q\, r)\theta(q_{\mathrm{max}}-|\vec{q}\,|)}{s-[\omega_i(q)+E_i(q)]^2 + i \varepsilon},
\end{equation}
where $E=\sqrt{s}$, $\omega_i=\sqrt{q^2+m_i^2}$ is the energy of the meson and $E_i$ the energy of the baryon.
\footnote{In Eq.~(13) of \cite{raquexiao} the $\theta(q_{\mathrm{max}}-|\vec{q}\,|)$ factor is missing. This is a typo and the right formula is given in Ref.~\cite{valcola}} It is interesting to see where the off-shell effects go. One of the $\theta(q_\mathrm{max}-|\vec{q}\,'|)$ factor of the $T$-matrix of Eq.~(\ref{eq:tm}) factorizes out of the integrals as $\theta(q_\mathrm{max}-p_i)$ and is inoperative since we always work with values of the momenta $p_i$ smaller than $q_\mathrm{max}$. The other $\theta(q_\mathrm{max}-|\vec{q}\,|)$ factor goes into the integral of $\tilde{G}$. This is interesting because it shows explicitly that the half off-shell amplitude is what enters the evaluation of the correlation function, as already stated in~\cite{epelulf, haiden}. 

If one had scattering data on $K\Lambda$, $K\Sigma$, etc, one could tune the parameter $q_\mathrm{max}$, assuming that the potential is known, and the chiral Lagrangians are a reliable source to produce this magnitude. Even then there is some possible trade-off between the strength of the potential and the value of $q_\mathrm{max}$. However, the value of the correlation functions is that they provide information on the interaction of pairs of particles for which a scattering experiment is not possible. This said, one has to admit that we do not have a precise knowledge on the value of $q_\mathrm{max}$ involved in the calculation of $\tilde{G}$, or in other words, we are faced with the uncertainties due to the off-shell extrapolation of the wave function, or equivalently the off-shell extrapolation of the scattering matrix. Yet, one has some feeling on reasonable extrapolations, because $q_\mathrm{max}$, as seen in Eq.~(\ref{eq:pot}), provides the range of the interaction in momentum space. Given the fact that the chiral Lagrangians can be obtained from vector exchange using the local hidden gauge approach~\cite{hidden1,hidden2,hidden4,hideko} (see an explicit derivation in Appendix A of Ref.~\cite{diastoledo}), the value of $q_\mathrm{max}$ should be of the order of $(600-1000)$~MeV, and we shall move in that range to see uncertainties in the correlation function from the off-shell extrapolation of the scattering matrix. Note that this goes in line with the comments in~Ref.~\cite{epelulf} when mentioning that the dominance of pion exchange in the NN interaction puts constraints on the freedom to go off-shell, which is investigated there by playing with different choices of the off-shell short range interactions. All this said, we show in the next section the changes obtained by going off-shell in the $T$-matrix. 
\section{Results}
\begin{figure}
 \centering
 \includegraphics[scale=0.5]{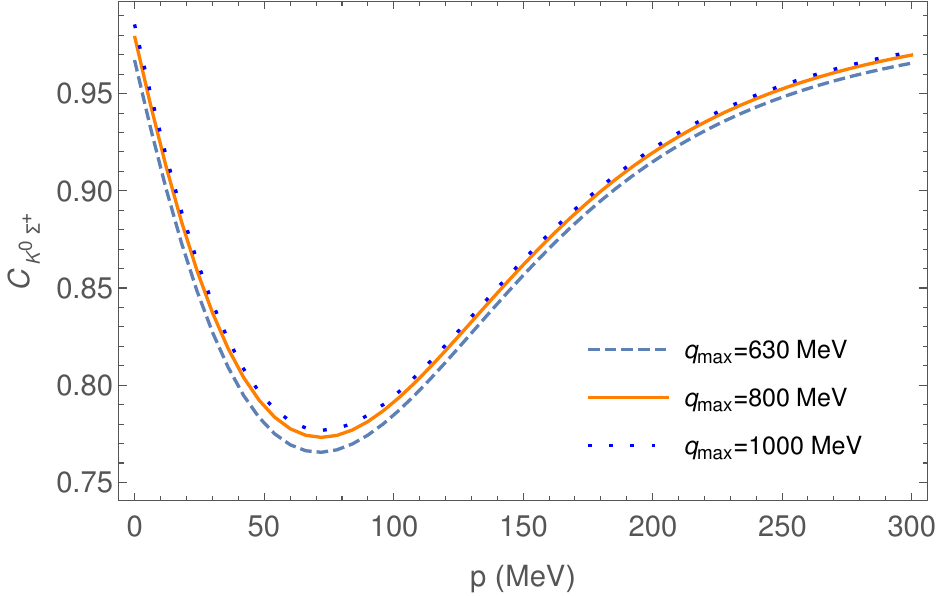}
 \caption{Correlation function of the $K^0\Sigma^+$ channel for different cut-offs and $R=1$~fm.}
 \label{fig:qmaxz}
\end{figure}

We take as a reference the correlation function of the channel $K^0\Sigma^+$ and start from the results of \cite{raquexiao} where the chiral Lagrangians were used to evaluate the transition potentials and a value of $q_\mathrm{max}=630$~MeV was used, the same one used in \cite{ramos} to get a good reproduction of data of $\bar{K}N$ and coupled channel. Then, we change $q_\mathrm{max}$ in the range indicated to quantify off-shell effects. The results are obtained by choosing the source radius $R=1$~fm. In Fig.~\ref{fig:qmaxz} we show the correlation function for values of $q_\mathrm{max}=630,800$ and $1000$~MeV. As we can see, the effects are moderate, but there is a dependence on $q_\mathrm{max}$. The differences between $q_\mathrm{max}=630$~MeV and $1000$~MeV are about $2$~\% at small values of $p$, and $1$\% at the bigger values. These differences are small, smaller than present experimental errors and nothing to worry about, unless one aims at obtaining information with this precision from future experiments~\cite{futurealice}. 

\begin{figure}
 \centering
 \includegraphics[scale=0.5]{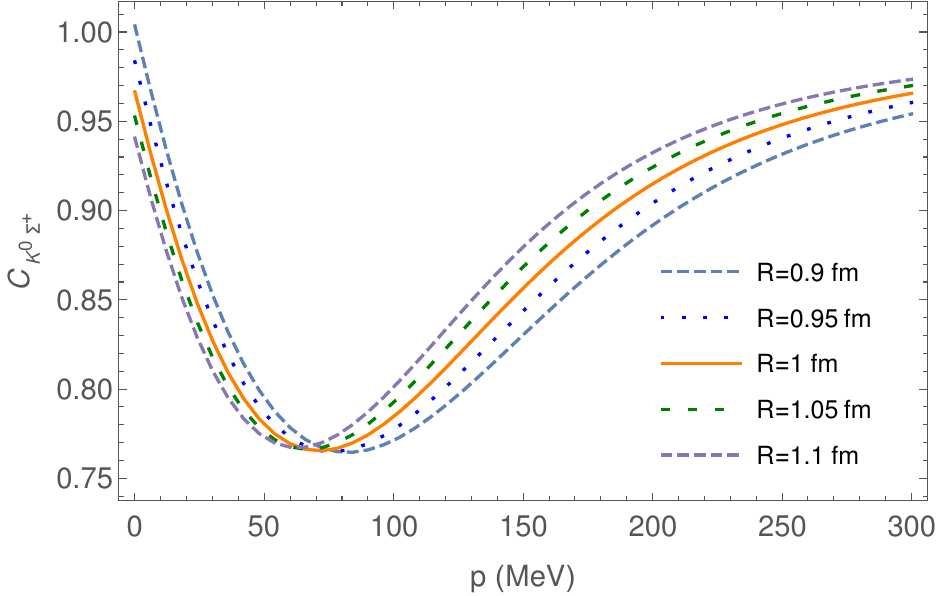}
 \caption{Correlation function of the $K^0\Sigma^+$ channel for different values of $R$ and $q_{\mathrm{max}}=630$~MeV.}
 \label{fig:dr}
\end{figure}

To put these results in perspective we show in Fig.~\ref{fig:dr} the results that we obtain changing the source size $R$ in $10$~\% up and down from the starting value of $R=1$~fm. 
As we can see, the changes obtained now are bigger, around $6$~\% change at small $p$ from $R=1.1$ fm to $R=0.9$~fm, also $6$~\% around $p=150$~MeV/c. For values of $p\sim 300$~MeV/c the differences are about $2$~\%. 

\begin{figure}
 \centering
 \includegraphics[scale=0.5]{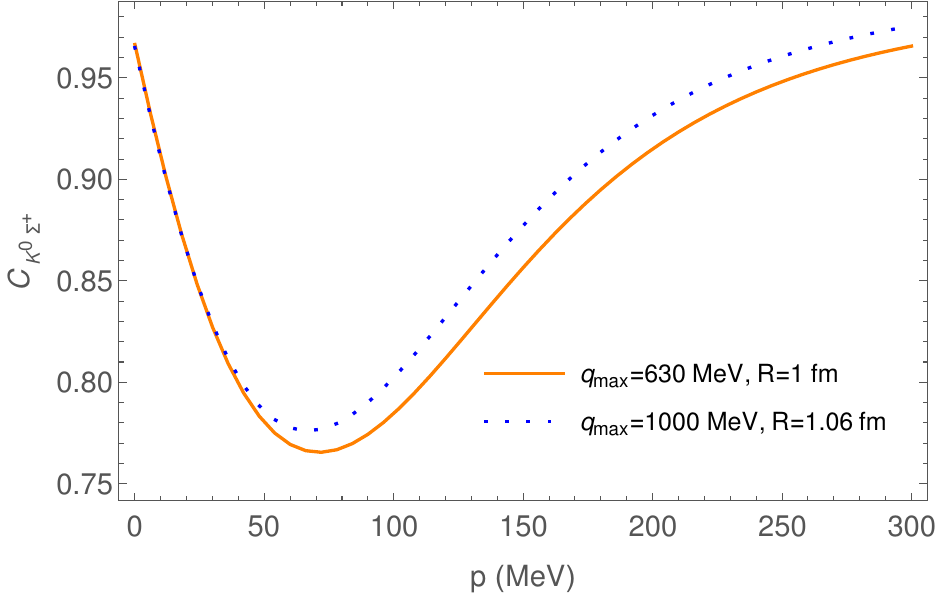}
 \caption{Correlation function of the $K^0\Sigma^+$ channel for different cut-offs/$R$'s.}
 \label{fig:qmax}
\end{figure}

\begin{figure}
 \centering
 \includegraphics[scale=0.5]{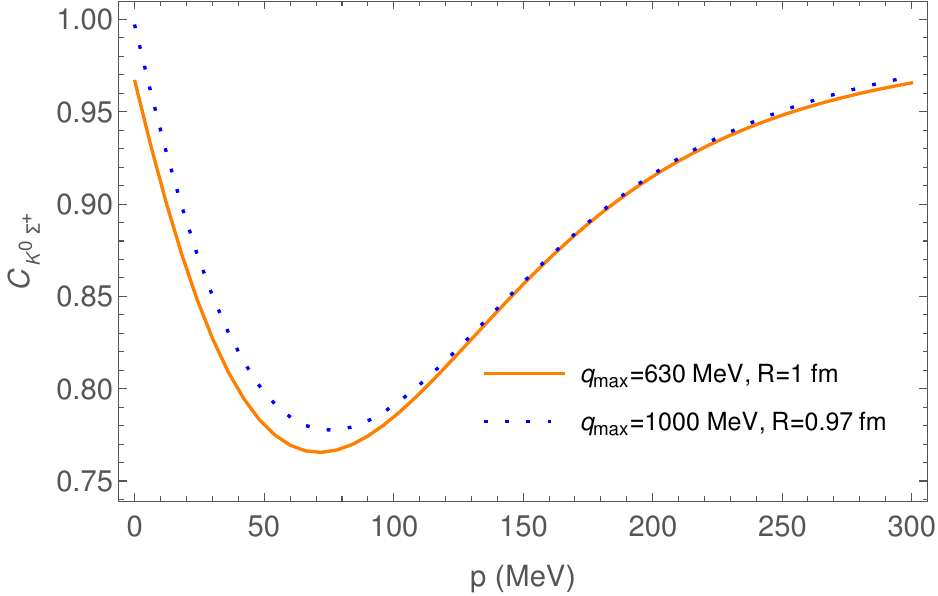}
 \caption{Correlation function of the $K^0\Sigma^+$ channel for different cut-offs/$R$'s.}
 \label{fig:c1mod}
\end{figure}

Changing the off-shell dependence of the $T$-matrix should be a consequence of some unitary transformation which changes the $T$-matrix keeping its on-shell value unchanged. As discussed in Ref.~\cite{epelulf} and the introduction, this unitary transformation should also be implemented in the source function to obtain the same correlation function. Then, we see if it is possible to change the source function with the same gaussian form of Eq.~(\ref{eq:sc}), such that we regain the original results for the correlation function before this unitary transformation is made. The results are shown in Figs.~\ref{fig:qmax} and \ref{fig:c1mod}. 

In Fig.~\ref{fig:qmax} we see that it is possible to change $R$ such that we obtain a perfect agreement with the original results up to about $p=50$~MeV/c, but then there are discrepancies at higher momenta. We learn three lessons: First, the unitary transformation on $S_{12}(r)$ corresponding to the unitary transformation required to go from $q_\mathrm{max}=630$~MeV to $q_\mathrm{max}=1000$~MeV, cannot be cast in terms of a Gaussian. Second, it is still possible to adopt this form if one restricts oneself to small values of $p$ below $50$~MeV/c. Third, by doing this, the changes at large $p\simeq 300$~MeV/c are now of the order of $1\%$, while before, the changes observed at low values of $p$ were of the order of $2$~\%. 

This is one of the conclusions of the paper: since one of the important informations that one obtains from correlation functions is the scattering length and effective range, which are encoded in the correlation function at small momenta $p$, the strategy to analyze the correlation data, when testing models for instance, is not to assume $R$ as a given data from experiment, and certainly not a universal value, but to fit $R$ to the data, together with other parameters of the theory. 

In Fig.~\ref{fig:c1mod} we show the results when we take an equivalent value of $R$ that fits the tail of the correlation function. In this case, we observe again that it is not possible to reproduce the full correlation function with just a change of the value of $R$, but we find a near perfect agreement from $p= 100$~MeV/c to $300$~MeV/c. The discrepancies at small $p$ are now of the order of $3$\%. While in both cases a change in $R$ improves restoring the changes induced by using a different off-shell extrapolation, the strategy of fitting the low energy part is better and minimizes the off-shell uncertainties. This suggest that it can be advisable to give more weight to the low momenta points of the correlation function in order to minimize the uncertainties tied to the off-shell ambiguities of the correlation function.

We have also looked at the correlation functions of the other coupled channels and the off-shell effects are similar for $K^+\Sigma^0$ and even smaller for the other channels. 

\begin{figure}
 \centering
 \includegraphics[scale=0.5]{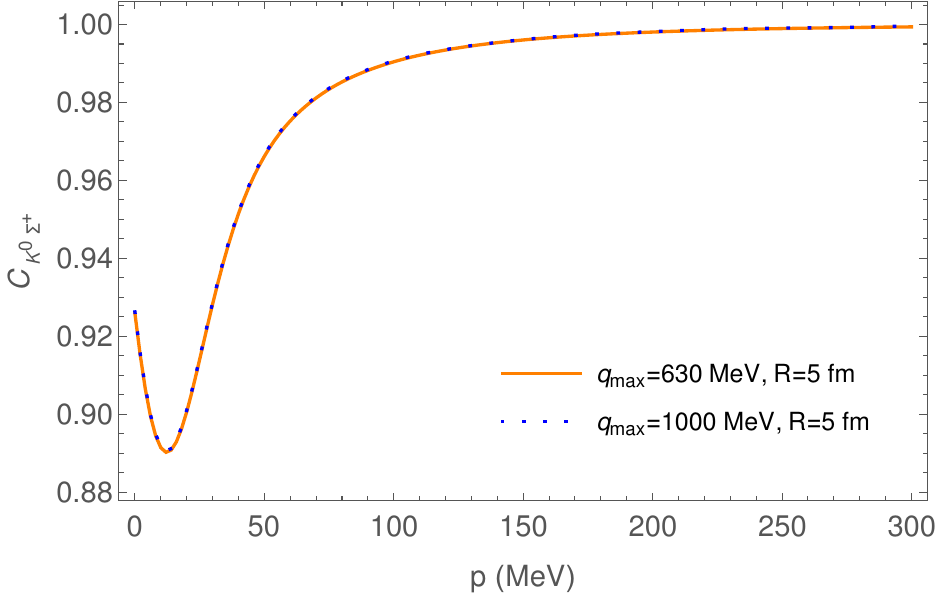}
 \caption{Correlation function of the $k^0\Sigma^+$ channel for $R=5$~fm and different cut-offs.}
 \label{fig:c1r}
\end{figure}
Finally, we show the off-shell effects but taking the source size $R=5$~fm. The results are shown in Fig.~\ref{fig:c1r} and, as we can see, the correlation functions for $q_\mathrm{max}=630$~MeV or $q_{\mathrm{max}}=1000$~MeV are indistinguishable. This trend, already advanced in \cite{epelulf}, is due to the role of the source function acting as a regulator through the $j_0(qr)$ factor in $\tilde{G}$ in Eq. (\ref{eq:G2}). This feature was discussed in detail in Ref.~\cite{juancorre} (see section IIC) and footnote 3 of that reference). 

In view of the negligible role of the off-shell effects for large $R$, it would look like that this situation would be ideal. However, we should note that the deviation from unity of the correlation function, which provides the information on the interaction of the particles, is much smaller in this case, and, it does not compensate the benefit of getting rid of the small uncertainties that we have reported for values of $R\simeq 1$~fm. 
\section {Conclusions}
We have studied the ambiguities in the correlation functions, due to off-shell extrapolations of the wave function, for the case of meson-baryon interaction, in particular the $K\Lambda$, $K\Sigma$, $\eta N$ coupled channels system. We use a formalism for the correlation functions based on the scattering matrices in momentum space, and the off-shell extrapolation is governed by changes in the regulator of the loop functions, a maximum momentum in the integrals, $q_{\mathrm{max}}$, entering the evaluation of the scattering matrices. We have shown that for realistic choices of the interaction, based on chiral dynamics, or extrapolations using the local hidden gauge approach, the uncertainties in the correlation functions are small, amounting to $2-3$ \%. Yet, the uncertainties could be bigger if other models are used. 

  We also found that changes in $q_{\mathrm{max}}$ can be compensated by a simultaneous change in the size of the source function, which also acts as a regulator of the loops, but only approximately, in a range of about $50-100$ MeV/c of momentum, not in the whole range of about $300$ MeV/c where the correlations functions are measured. One has the choice of fitting the upper or the lower part of the spectrum of momenta, but we find more rewarding to fit the lower part of the spectrum from where the scattering length and effective range are obtained.   With all these findings we come to the following recommendations when fitting a model to the data. 
  \begin{itemize}
\item[1)] Do not take the source size $R$ for granted. It is not universal. Then take R as one more parameter of the theory to fit the data. 
\item[2)] Since the correlation functions basically provide the scattering length and effective range, which are tied to the low momentum part of the spectrum, it is convenient to put more weight in the fit in this region. It does not pay to put equal or more weight on the upper part of the spectrum, because once the low energy part is well fitted, there are remaining off-shell ambiguities in the upper part of the spectrum. 
\item[3)] An alternative to testing models is to use a model independent method to determine the observables from the correlation functions: scattering lengths, effective ranges and the existence of possible bound states below thresholds. This method does not imply any model, but parametrizes the potentials, and uses these parameters, together with the regulator $q_{\mathrm{max}}$, and $R$ as free parameters. It has been proved in a series of papers that this inverse problem approach is very efficient \cite{ikenoinv,alberinv,daiinv,raquexiao,haipenginv}. The approach is accompanied by a resampling method to determine the uncertainties of the observables obtained. Many fits are done by changing randomly the centroids of the data within the experimental errors, and the parameters of the theory are calculated each time, and from them the observables. This is most indicated since there are correlations between the parameters, and the resampling method takes care of it. In each of the fits of the resampling the parameters obtained are different, but the observables are stable and their variation in the different fits determines their errors. 
\end{itemize}
  In summary, we concluded that in the studied case of the meson baryon interaction, and one thinks that this would be the case in other similar problems, the uncertainties in the correlation functions due to off-shell ambiguities are small, of the order of $2-3$ \%, when using realistic models based on chiral dynamics. However, these uncertainties can be bigger if one uses other models. In view of this, or in any case if one wishes to have very high precision in the observables obtained, we give some recommendations on how to proceed when analyzing correlation functions to extract scattering observables from them to minimize uncertainties from off-shell ambiguities in the wave function.

\section{Acknowledgments}

We thank to A. Feijoo for a careful reading of the manuscript. 
R. M. acknowledges support from ESGENT program (ESGENT/018/2024) and the PROMETEU program (CIPROM/2023/59), of
the Generalitat Valenciana, and also from the Spanish Ministerio de Economia y Competitividad (MINECO) and European Union (NextGenerationEU/PRTR) by the grants with Ref. CNS2022-13614, and Ref. PID2023-147458NB-C21.

\bibliography{biblio}

\end{document}